\documentclass[a4paper]{lipics-v2018}

\usepackage{hyperref}
\usepackage{enumerate}
\usepackage{proof}
\usepackage{amsmath}
\usepackage{stmaryrd}
\usepackage{alltt}
\usepackage{listings}
\usepackage{color}
\usepackage{lstcoq}
\usepackage{listings}
\usepackage{lstlinebgrd}
\usepackage{pdflscape}
\usepackage{rotating}

\newcommand{\BLUE}[1]{{\color{blue} #1}}

\usepackage[all]{xy}
\usepackage{amssymb}
\usepackage{wrapfig}

\newif\iflongversion

\longversiontrue

\newcommand{\Impl}{\supset}

\newcommand {\Type}      {{\mathsf {Type}}}

\newcommand {\ala}   {{\textit{\`a la}}}
\newcommand {\ie}    {{\textit{i}.\textit{e}.}}

\newcommand {\eg}    {{\textit{e}.\textit{g}.}}

\newcommand {\etalt} {{\textit{et al.}}}
\newcommand {\adhoc} {{\textit{ad hoc}}}

\newcommand{\red}{\longrightarrow}

\newcommand {\at}          {\,}

\newcommand {\atr}        {\hspace{.5mm}^{_{_{_\textsf{r}}}}}

\newcommand {\of}          {{:}}

\newcommand{\D}{\ensuremath{\Delta}}

\newcommand{\eqdef}{\stackrel{\mathit{def}}{=}}

\newcommand{\s}{\ensuremath{\sigma}}
\renewcommand{\t}{\ensuremath{\tau}}
\renewcommand{\r}{\ensuremath{\rho}}

\newcommand{\spair}[2]{\ensuremath{\langle {#1}\,\,  ,\,{#2} \rangle }}

\newcommand{\ssum}[2]{\ensuremath{[ {#1}\,\, ,\,{#2} ]}}

\newcommand{\inll}[1]{{in}^{#1}_l\,}
\newcommand{\inrr}[1]{{in}^{#1}_r\,}

\newcommand{\inl}{{in}_l\,}
\newcommand{\inr}{{in}_r\,}
\newcommand{\ini}{{in}_i\,}

\newcommand{\prl}{{pr}_{\!l}\,}
\newcommand{\prr}{{pr}_{\!r}\,}
\newcommand{\pri}{{pr}_{\!i}\,}

\newcommand{\lambdar}{\lambda^{\tiny \! \mathsf r} }
\newcommand{\Pir}{\Pi^{\tiny \mathsf r}}

\newcommand{\erase}[1]{\mid\!{#1}\!\mid}
\newcommand{\erasetype}[1]{\mid\!\mid\!{#1}\!\mid\!\mid}

\newcommand{\essence}[1]{{\wr}{\,#1\,}{\wr}}

\newcommand{\sigmadash}{\vdash_\Sigma}

\newcommand{\oksig}[1]{#1\text{ sig}}
\newcommand{\emptysig}{\langle\rangle} 
\newcommand{\emptyctx}{\langle\rangle}
\newcommand{\dom}[1]{\text{dom}(#1)}

\newcommand{\LF}{LF}
\newcommand{\DLF}{\mbox{\rm LF$_\Delta$}}
\newcommand{\B}{{\mathcal B}}

\title{The $\Delta$-framework}

\titlerunning{The $\Delta$-framework}

\author{}{}{}{}{}

\author{Furio Honsell$^1$ \quad Luigi Liquori$^2$  \quad Claude Stolze$^2$ \quad Ivan Scagnetto$^1$}
{\centering($1$) University of Udine \quad $(2)$ Universit\'e C\^ote d'Azur, INRIA}{}{}{}
\authorrunning{F. Honsell, L. Liquori, C. Stolze and I. Scagnetto}

\nolinenumbers

\begin{document}

\maketitle

\begin{abstract}
We introduce the $\Delta$-framework, \DLF,  a dependent type theory based on the Edinburgh Logical Framework LF, extended with the {\em strong  proof-functional connectives}, \ie\ strong intersection, minimal relevant implication and strong union. Strong proof-functional connectives take into account the shape of logical proofs, thus reflecting polymorphic features of proofs in formul\ae. This is in contrast to classical or intuitionistic connectives where the meaning of a compound formula depends only on the truth value or the provability of its subformul\ae.
Our framework encompasses a wide range of type disciplines. Moreover, since relevant implication permits to express subtyping, \DLF\ subsumes also Pfenning's refinement types. We discuss the design decisions which have led us to the formulation of \DLF, study its metatheory, and provide various examples of applications. Our strong proof-functional type theory can be plugged in existing common proof assistants.
\end{abstract}

\maketitle


\vspace{-1cm}\section{Introduction}
\noindent
This paper provides a unifying framework for two hitherto unreconciled understandings of types: \ie\ types-as-predicates \ala\ Curry and types-as-propositions (sets) \ala\ Church. The key to our unification consists in introducing {\em strong proof-functional connectives} \cite{pott80,BDdL,BM94} in a dependent type theory such as the Edinburgh Logical Framework (LF) \cite{LF}.
Both Logical Frameworks and Proof-Functional Logics consider proofs as first class citizens, albeit differently.
Strong proof-functional connectives take seriously into account the shape of logical proofs, thus allowing for polymorphic features of proofs to be made explicit in formul\ae. They provide a finer semantics than classical/intuitionistic connectives, where the meaning of a compound formula depends only on the {\em truth value} or the {\em provability} of its subformul\ae. However, existing approaches to strong proof-functional connectives are all quite idiosyncratic in mentioning proofs.
Logical Frameworks, on the other hand, provide a uniform approach to proof terms in object logics, but they have not fully capitalized on subtyping.

This situation calls for a natural combination of the two understandings of types, which should benefit both worlds. On the side of Logical Frameworks, the expressive power of the metalanguage would be enhanced thus allowing for shallower encodings of logics, a more principled use of subtypes \cite{Refine93}, and new possibilities for formal reasoning in existing interactive theorem provers. On the side of type disciplines for programming languages, a principled framework for proofs would be provided, thus supporting a uniform approach to ``proof reuse'' practices based on type theory \cite{pier91b,caplan,felty94,boite,barthe-pons}.

Therefore, in this paper, we extend LF with the connectives of {\em strong intersection}, {\em strong union}, and {\em minimal relevant implication} of Proof-Functional Logics \cite{pott80,BDdL,BM94}. We call this extension the $\Delta$-framework
(\DLF), since it builds on the $\Delta$-calculus introduced in \cite{LS18,APLAS16}.
Moreover, we illustrate by way of examples, that \DLF\ subsumes many expressive type disciplines in the literature \cite{Refine93,BDdL,BM94,pier91b,caplan}.

It is not immediate to extend the judgments-as-type, Curry-Howard paradigm to logics supporting strong proof-functional connectives, since these connectives need to compare the shapes of derivations and do not just take into account the provability of propositions, \ie\ the inhabitation of the corresponding type. In order to capture successfully strong logical connectives such as $\cap$ or $\cup$, we need to be able to express the rules:

\noindent $
\begin{array}{c@{\qquad} c}
\infer[(\cap I)]
{A \cap B}
{{\mathcal D}_1 : A \quad {\mathcal D}_2 : B \quad {\mathcal D}_1 \equiv {\mathcal D}_2}
&
\infer[(\cup E)]
{C}
{{\mathcal D}_1 : A \supset C \quad {\mathcal D}_2 : B \supset C \quad A \cup B \quad {\mathcal D}_1 \equiv {\mathcal D}_2 }
\end{array}
$
where $\equiv$ is a suitable equivalence between logical proofs.
Notice that the above rules suggest immediately intriguing applications in polymorphic constructions, \ie\ the same evidence can be used as a proof for different statements.
Pottinger \cite{pott80} was the first to study the strong connective $\cap$. He contrasted it to the intuitionistic connective $\wedge$ as follows: \emph{``The intuitive meaning of $\cap$ can be explained by saying that to assert $A \cap B$ is to assert that one has a reason for asserting $A$ which is also a reason for asserting $B$ ... (while) ... to assert $A \wedge B$ is to assert that one has a pair of reasons, the first of which is a reason for asserting $A$ and the second of which is a reason for asserting $B$''}.
A logical theorem involving intuitionistic conjunction which does not hold for strong conjunction is $(A\supset A) \wedge (A \supset B \supset A)$, otherwise there should exist a closed $\lambda$-term having simultaneously both one and two abstractions.
Lopez-Escobar \cite{Lopez-Escobar85} and Mints \cite{Mints89} investigated extensively logics featuring both strong and intuitionistic connectives especially in the context of {\em realizability} interpretations.

Dually, it is in the $\cup$-elimination rule that proof equality needs to be checked. Following Pottinger, we could say that {\em asserting $(A \cup B) \Impl C$ is to assert that one has a reason for $(A \cup B) \Impl C$, which is also a reason to assert $A \Impl C$ and $B \Impl C$.} The two connectives differ since the intuitionistic theorem $((A \supset B) \vee B) \supset A \supset B$ is not derivable for $\cup$, otherwise there would exist a term which behaves both as {\bf I} and as {\bf K}.

Following \cite{BM94}, \emph{Strong} (or \emph{Minimal Relevant}) \emph{Implication}, $\supset_r$, can be viewed as a special case of implication whose related function space is the simplest possible one, namely the one containing only the identity function. The operators $\supset$ and $\supset_r$ differ, since $A \supset_r B \supset_r A$ is not derivable. Relevant implication allows for a natural introduction of subtyping, in that $A \supset_r B$ morally means $A \leqslant B $. Relevant implication amounts to a notion of ``proof-reuse''. 
Combining the remarks in \cite{BM94,BDdL}, minimal relevant implication, strong intersection and strong union correspond respectively to the implication, conjunction and disjunction operators of Meyer and Routley's Minimal Relevant Logic $B^+$ \cite{RM72}.

\begin{figure}[t]
$$
\begin{array}[t]{r@{\qquad \qquad}r}
\multicolumn{2}{c}
{\infer[(\cap I)]
 {B \vdash M : \s \cap \t}
 {B \vdash M : \s &
 B \vdash M : \t}
\quad
\infer[(\cap E_i)]
 {B \vdash M : \s}
 {B \vdash M : \s \cap \t }
\quad
\infer[(\cap E_r)]
 {B \vdash M : \t}
 {B \vdash M : \s \cap \t}
 }
 \\[2mm]
 \infer[(\cup I_l)]
 {B \vdash M : \s \cup \t}
 {B \vdash M : \s }
 &
 \infer[(\cup I_r)]
 {B \vdash M : \s \cup \t}
 {B \vdash M : \t }
 \\[2mm]
 \multicolumn{2}{r}
 {\infer[(\cup E)]
 {B \vdash M[N/x] : \r}
 {
 B,x\of\s \vdash M : \r &
 B,x\of\t \vdash M : \r &
 B \vdash N : \s \cup \t
 } \quad \infer[(Sub)]
{B \vdash M : \t}
{B \vdash M : \s & \s \leq \t}
 }
 \\[2mm]
 \multicolumn{2}{c}
 {\infer[(Var)]
{B \vdash x : \s}
{x \of \s \in B}
\quad
\infer[(App)]
{B \vdash M \at N : \t}
{B \vdash M : \s \to \t & B \vdash N : \s}
\quad
\infer[(Abs)]
{B \vdash \lambda x.M : \s \to \t}
{B,x\of\s \vdash M:\t}}
\end{array}
$$

$$
\begin{array}[t]{l@{\quad}l@{\qquad} r@{\quad } l}
(1) & \sigma \leqslant \sigma \cap \sigma
 &
(8) & \sigma_1 \leqslant \sigma_2, \tau_1 \leqslant \tau_2 \Rightarrow \sigma_1 \cup \tau_1 \leqslant \sigma_2 \cup \tau_2\\[2mm]
(2) & \sigma \cup \sigma \leqslant \sigma
 &
(9) & \sigma \leqslant \tau, \tau \leqslant \rho \Rightarrow \sigma \leqslant \rho\\[2mm]
(3) & \sigma \cap \tau \leqslant \sigma, \sigma \cap \tau \leqslant \tau
&
(10) & \sigma \cap (\tau \cup \rho) \leqslant (\sigma \cap \tau) \cup (\sigma \cap \rho)\\[2mm]
(4) & \sigma \leqslant \sigma \cup \tau, \tau \leqslant \sigma \cup \tau
&
(11) & (\sigma \rightarrow \tau) \cap (\sigma \rightarrow \rho) \leqslant \sigma \rightarrow (\tau \cap \rho)\\[2mm]
(5) & \sigma \leqslant \omega
&
(12) & (\sigma \rightarrow \rho) \cap (\tau \rightarrow \rho) \leqslant (\sigma \cup \tau) \rightarrow \rho\\[2mm]
(6) & \sigma \leqslant \sigma
&
(13) & \omega \leqslant \omega \rightarrow \omega\\[2mm]
(7) & \sigma_1 \leqslant \sigma_2, \tau_1 \leqslant \tau_2 \Rightarrow
 \sigma_1 \cap \tau_1 \leqslant \sigma_2 \cap \tau_2
&
(14) & \sigma_2 \leqslant \sigma_1, \tau_1 \leqslant \tau_2 \Rightarrow
\sigma_1 \rightarrow \tau_1 \leqslant \sigma_2 \rightarrow \tau_2
\end{array}
$$
\caption{The type assignment system $\B$ of \cite{BDdL} and the subtype theory $\Xi$}
 \label{curryandchurch}
\end{figure}

Strong connectives arise naturally in investigating the proposi\-tions-as-types analogy for intersection and union type assignment systems. Intersection types were introduced by Coppo, Dezani \etalt\ in the late 70's \cite{CD78,CDS,CDV,BCD} to support a form of \adhoc\ polymorphism, for untyped $\lambda$-calculi, \ala\ Curry. Intersection types were used originally as an (undecidable) type assignment system for pure $\lambda$-calculi, \ie\ for finitary descriptions of denotational semantics \cite{HL-TCS_212}. This line of research was later explored by Abramsky \cite{DTLF} in a full-fledged Stone duality.
Union types were introduced semantically, by MacQueen, Plotkin, and Sethi \cite{macqueen,BDdL}. In \cite{BDdL} strong intersection, union and subtyping were thoroughly studied in the context of type-assignment systems, see Figure~\ref{curryandchurch}. A classical example of the expressiveness of union types is due to Pierce \cite{pier91b}: without union types, the best information we can get for
\begin{wrapfigure}{l}{0.65\textwidth}
\vspace{-3mm}
$
\begin{array}{rcl}
\mbox{Test}
&\eqdef &
\mbox{if } b
\mbox{ then } 1 \mbox{ else} -\!1 : Pos \cup Neg
\\
\mbox{Is}\_0
& : &
(Neg \to F) \cap
(Zero \to T) \cap
(Pos \to F)
\\
(\mbox{Is}\_0 \mbox{ Test}) &:& F
\end{array}
$
 \vspace{-3mm}
\end{wrapfigure}
$(\mbox{Is}\_0 \at \mbox{Test})$ is a boolean type.
Designing a $\lambda$-calculus \ala\ Church with intersection and union types is problematic.
The usual approach of simply adding types to binders does not work, as shown in Figure \ref{polyid}. 
\begin{wrapfigure}{l}{0.55\textwidth}
\vspace{-3mm}
$
\infer[(\cap I)]{\vdash \lambda x \of {\color{red}{???}}. x \of (\s \to \s) \cap (\t \to \t)}{
 \infer[({\to}I)]{\vdash \lambda x \of {\color{red}{\s}}. x \of \s \to \s}{
 \infer[(V\!ar)]{x \of \s \vdash x \of \s}{}
 } \quad &
 \infer[({\to}I)]{\vdash \lambda x \of {\color{red}{\t}}. x \of \t \to \t}{
 \infer[(V\!ar)]{x \of \t \vdash x \of \t}{}
 }
 }
$
\caption{Polymorphic identity}
\label{polyid}
\vspace{-3mm}
\end{wrapfigure}
Same difficulties can be found with union types. Intersection and union type disciplines started to be investigated in a explicitly typed programming language settings \ala\ Church, much later by Reynolds and Pierce \cite{reyn88,pier91b}, Wells \etalt\ \cite{wells2,wells3}, Liquori \etalt\ \cite{LR07,Dougherty-Liquori}, Frisch \etalt\ \cite{Frisch08} and Dunfield \cite{Dunfield}. From a logical point of view, there are many proposals to find a suitable logics to fit intersection: among them we cite \cite{Refine93,venneri94,roncrove01,miquel01,CLV,BVB,pimronrov12}, and previous papers by the authors \cite{APLAS16,TTCS17} and a type checker implementation  \cite{LFMTP17}. In \cite{APLAS16}, two of the present authors proposed the $\Delta$-calculus as a typed $\lambda$-calculus \ala\ Church corresponding to the type assignment system \ala\ Curry with intersection and union
but without $\omega$. The relation between Church-style and Curry-style $\lambda$-calculi was expressed using an \emph{essence} function, denoted by $\essence{-}$, that intuitively erases all the type information in terms (the full definition is shown in Figure \ref{ESSENCE}). 


\noindent  The \DLF, introduced in this paper extends \cite{LS18} with union types, 
dependent types and minimal relevant implication. The novelty of \DLF\ in the context of Logical Frameworks, lies in the full-fledged use of strong proof-functional connectives, which to our knowledge has never been explored before. Clearly, all $\Delta$-terms have a computational counterpart.


Pfenning's work on Refinement Types \cite{Refine93} pioneered an extension of the Edinburgh Logical Framework with subtyping and intersection types. His approach capitalises on a tame and essentially \adhoc\ notion of subtyping, but the logical strength of that system does not go beyond the \LF\ (\ie\ simple types). The logical power of \DLF\ allows to type all strongly normalizing terms. Furthermore, subtyping in \DLF\ arises naturally as a derived notion from the more fundamental concept of minimal relevant implication, as illustrated in Section \ref{sec:DLF}.

Miquel \cite{miquel01} discusses an extension of the Calculus of Constructions with implicit typing, which subsumes a kind of proof-functional intersection. His approach has opposite motivations to ours. While \DLF\ provides a Church-style version of Curry-style type assignment systems, Miquel's Implicit Calculus of Constructions encompasses some features of Curry-style systems in an otherwise Church-style Calculus of Constructions. In \DLF\ we can discuss also \adhoc\ polymorphism, while in the Implicit Calculus only structural polymorphism is encoded. Indeed, he cannot assign the type $((\sigma \cap \tau)\rightarrow \sigma)\cap (\rho \rightarrow \rho))$ to the identity $\lambda x.x$ \cite{private}. Kopylov~\cite{Kopylov2003} adds a dependent intersection type constructor $x\of A\cap B[x]$ to NuPRL, allowing the resulting system to support dependent records (which are a very useful data structure to encode mathematics). The implicit product-type of Miquel, together with the dependent intersection type of Kopylov, and a suitable equality-type is used by Stump~\cite{Stump2018} to  enrich the impredicative second-order system $\lambda P2$, in order to derive induction.

In order to achieve our goals, we could have carried out simply the encoding of \DLF\ in \LF. But, due to the side-conditions characterizing proof-functional connectives, this would have be achieved only through a deep encoding. 
As an example of this, in Figure \ref{BDDLINCOQ}, we give an encoding of a subsystem of \cite{BDdL}, where subtyping has been simulated using relevant arrows. This encoding illustrates the expressive power of \LF\ in treating proofs as first-class citizens, and it was also a source of inspiration for \DLF.

All the examples discussed in this paper have been checked by an experimental proof development environment for \DLF\ \cite{LFMTP17} (see {\color{blue} \href{https://github.com/cstolze/Bull}{Bull}} and {\color{blue} \href{https://github.com/cstolze/Bull-Subtyping}{Bull-Subtyping}} in \cite{BULL}).
 
\noindent {\bf Synopsis.} In Section \ref{sec:DLF}, we introduce \DLF\ and outline its metatheory, together with a discussion of the main design decisions. In Section \ref{sec:examples}, we provide the motivating examples. In Section \ref{sec:impl}, we outline the details of the implementation and future work.

%

\begin{figure}[t]
$$\!\!\!\!\!\!\!\!
\begin{array}{rcl@{\quad}l}
\multicolumn{4}{l}
{\rm Kinds}
\\[1mm]
K & ::= & \Type \mid \Pi x \of \s.K 
& \mbox{as in LF}
\\[2mm]
\multicolumn{4}{l}
{\rm Families}
\\[1mm]
\s,\t & ::= &
a \mid \Pi x \of \s.\t
\mid \s \at \Delta
\mid
& \mbox{as in LF}
\\[1mm]
&&
\s\rightarrow^{\sf r}\t \mid
& \mbox{relevant family}
\\[1mm]
&&
\s \cap \tau\mid
& \mbox{intersection family}
\\[1mm]
&&
\s \cup \tau
& \mbox{union family}
\\[4mm]
\end{array}
\hspace{-10mm}
\begin{array}{rcl@{\quad}l}
\multicolumn{4}{l}
{\rm Objects}
\\[1mm]
\Delta & ::= & c \mid x \mid
\lambda x \of \s.\Delta \mid \Delta \at \Delta \mid
& \mbox{as in LF}
\\[1mm]
&&
\lambdar x \of\s.\Delta \mid
& \mbox{relevant abstraction}
\\[1mm]
&&
\Delta \atr \Delta \mid
& \mbox{relevant application}
\\[1mm]
&&
\spair{\Delta}{\Delta} \mid
& \mbox{intersection objects}
\\[1mm]
&&
\ssum{\Delta}{\Delta} \mid
& \mbox{union objects}
\\[1mm]
&&
\prl{\Delta} \mid \prr{\Delta} \mid
& \mbox{projections objects}
\\[1mm]
&&
\inll{\sigma}{\Delta} \mid \inrr{\sigma}{\Delta}
& \mbox{injections objects}
\\[1mm]
\end{array}
$$
\caption{The syntax of the $\Delta$-framework}
\label{SYN}
\end{figure}

\section{The $\Delta$-framework: LF with proof-functional operators}\label{sec:DLF}
The syntax of \DLF\ pseudo-terms is given in Figure~\ref{SYN}.
%
For the sake of simplicity, we suppose that $\alpha$-convertible terms are equal. Signatures and contexts are defined as finite sequence of declarations, like in LF.
Observe that we could formulate \DLF\ in the style of \cite{harper-licata}, using only canonical forms and without reductions, but we prefer to use the standard \LF\ format to support better intuition.
There are three proof-functional objects, namely strong conjunction (typed with $\s \cap \t$) with two corresponding projections, strong disjunction (typed with $\s \cup \t$) with two corresponding injections, and strong (or relevant) $\lambda$-abstraction (typed with $\rightarrow^{\sf r}$). Indeed, a relevant implication is not a dependent one because the essence of the inhabitants of type $\s\rightarrow^{\sf r}\t$ is essentially the identity function as enforced in the typing rules. Note that injections $\ini$ need to be decorated with the injected type $\s$ in order to ensure the unicity of typing.

\begin{figure}[t]
$$
\begin{array}{rcl}
\essence{\spair{\Delta_1}{\Delta_2}}
&\eqdef &
\essence{\Delta_1}
\\[2mm]
\essence{\lambdar x \of \s.\Delta}
& \eqdef &
\lambda x. \essence{\Delta}
\\[2mm]
\essence{\lambda x \of \s. \Delta}
& \eqdef &
\lambda x. \essence{\Delta}
\end{array}
\hspace{1cm}
\begin{array}{rcl}
\essence{\ssum{\Delta_1}{\Delta_2}}
&\eqdef &
\essence{\Delta_1}
\\[2mm]
\essence{\Delta_1 \at \Delta_2}
& \eqdef &
\essence{\Delta_1} \atr \essence{\Delta_2}
\\[2mm]
\essence{\Delta_1 \at \Delta_2}
& \eqdef &
\essence{\Delta_2}
\end{array}
\hspace{1cm}
\begin{array}{rcl}
\essence{\pri{\Delta}}
& \eqdef &
\essence{\Delta}
\\[2mm]
\essence{\ini{\Delta}}
& \eqdef &
\essence{\Delta}
\\[2mm]
\essence{c}
& \eqdef & c
\\[0mm]
\essence{x}
& \eqdef & x
\end{array}
$$
\caption{The essence function}
\label{ESSENCE}
\end{figure}

We need to generalize the notion of \emph{essence}, introduced in \cite{APLAS16,TTCS17} to syntactically connect pure $\lambda$-terms (denoted by $M$) and type annotated \DLF\ terms (denoted by $\Delta$). The essence function compositionally erases all type annotations, see Figure \ref{ESSENCE}.

One could argue that the choice of $\D_1$ in the definition of strong pairs/co-pairs is arbitrary and could have been replaced with $\D_2$: however, the typing rules will ensure that, if $\spair{\D_1}{\D_2}$ (resp. $\essence{\ssum{\D_1}{\D_2}}$) is typable, then we have that $\essence{\D_1} \mathrel{=_\eta} \essence{\D_2}$. Thus, strong pairs/co-pairs are constrained.
%
The rule for the essence of a relevant application is justified by the fact that the operator amounts to just a type decoration.

\begin{figure*}[t]
$$
\begin{array}{rll}
(\lambda x\of\s.\Delta_1) \at \Delta_2 & \red_\beta & \Delta_1[\Delta_2/x]
\\[2mm]
\prl \spair{\Delta_1}{\Delta_2} & \red_{\prl} & \Delta_1
\\[2mm]
\prr \spair{\Delta_1}{\Delta_2} & \red_{\prr} & \Delta_2
\\[2mm]
\ssum{\Delta_1}{\Delta_2} \at \inll{\s} \Delta_3 & \red_{\inl} & \Delta_1 \at \Delta_3
\\[2mm]
\ssum{\Delta_1}{\Delta_2} \at \inrr{\s} \Delta_3 & \red_{\inr} & \Delta_2 \at \Delta_3
\\[2mm]
(\lambdar x\of\s.{\Delta_1})\atr \Delta_2 & \red_{\beta r} & \Delta_1[\Delta_2/x]
\end{array}
\quad
\begin{array}{rll}

\multicolumn{3}{c}
{
\infer[(Congr_\cap)]{\spair{\Delta_1}{\Delta_2} \mathrel{\to_{\Delta}} \spair{\Delta'_1}{\Delta'_2}}
 {\Delta_1 \mathrel{\to_{\Delta}} \Delta'_1 & \Delta_2 \mathrel{\to_{\Delta}} \Delta'_2 & \essence{\Delta'_1}\equiv\essence{\Delta'_2}}
}
 \\[2mm]
\multicolumn{3}{c}
{
\infer[(Congr_\cup)]{\ssum{\Delta_1}{\Delta_2} \mathrel{\to_{\Delta}} \ssum{\Delta_1'}{\Delta_2'}}
 {\Delta_1 \mathrel{\to_{\Delta}} \Delta_1' & \Delta_2 \mathrel{\to_{\Delta}} \Delta_2' & \essence{\Delta'_1}\equiv\essence{\Delta'_2}}
}
\end{array}
$$

\caption{The reduction semantics}
\label{fig:congruence}
\end{figure*}

The six basic reductions for \DLF\ objects appear on the left in Figure \ref{fig:congruence}. Congruence rules are as usual, except for the two cases dealing with pairs and co-pairs which appear on the right of Figure \ref{fig:congruence}. Here redexes need to be reduced ``in parallel'' in order to preserve identity of essences in the components. We denote by $=_{\Delta}$ the symmetric, reflexive, and transitive closure of $\to_{\Delta}$, \ie\ the compatible closure of the reduction induced by the first six rules on the left in Figure \ref{fig:congruence}, with the addition of the last two congruence rules in the same figure.
In order to make this definition truly functional as well as to be able to prove a simple subject reduction result, we need to constrain pairs and co-pairs, \ie\ objects of the form $\spair{\Delta_i}{\Delta_j}$ and $\ssum{\Delta_i}{\Delta_j}$ to have congruent components up-to erasure of type annotations. This is achieved by imposing $\essence{\Delta_i} \equiv  \essence{\Delta_j}$ in both constructs. We will therefore assume that such pairs and co-pairs are simply not well defined terms, if the components have a different ``infrastructure''. The effects of this choice are reflected in the congruence rules in the reduction relation, in order to ensure that reductions can only be carried out in parallel along the two components.

The restriction on reductions in pairs/co-pairs and the new constructs do not cause any problems in showing that $\rightarrow_\Delta$ is locally confluent:

\begin{theorem}[Local confluence] \label{prop:localconfluence}\hfill\\
The reduction relation on well-formed \DLF-terms is locally confluent.
\end{theorem}
The extended type theory \DLF\ is a formal system for deriving judgements of the forms:
$$
\begin{array}{rll@{\qquad}l}
& \vdash & \Sigma & \Sigma ~{\rm is~a~valid~signature}
\\[1mm]
& \sigmadash & \Gamma & \Gamma~{\rm is~a~valid~context~in}~\Sigma
\\[1mm]
\Gamma & \sigmadash & K & K~{\rm is~a~kind~in~\Gamma~and~\Sigma}
\end{array}
\quad
\begin{array}[b]{rll@{\qquad}l}
\Gamma & \sigmadash & \s : K & \sigma~{\rm has~kind~K~in~\Gamma~and~\Sigma}
\\[1mm]
\Gamma & \sigmadash & \Delta :\s & \Delta~{\rm has~type~\sigma~in~\Gamma~and~\Sigma}
\end{array}
$$
\begin{figure}[t!]
\begin{center}
\begin{math}
\begin{array}{l@{\qquad}r}
\multicolumn{2}{l}{\mbox{Valid Objects}}
\\[4mm]
\infer[(Const)]
{\Gamma \sigmadash c : \s}
{\sigmadash \Gamma & c \of \s \in \Sigma}
&
\infer[(V\!ar)]
{\Gamma \sigmadash x : \s}
{\sigmadash \Gamma & x \of \s \in \Gamma}
\\[4mm]
\infer[(\Pi I)]
{\Gamma \sigmadash \lambda x \of \s.\Delta : \Pi x \of \s. \t}
{\Gamma,x \of \s \sigmadash \Delta : \t}
&
\infer[(\Pi E)]
{\Gamma \sigmadash \Delta_1\at \Delta_2 : \t[\Delta_2/ x ]}
{\Gamma \sigmadash \Delta_1 : \Pi x \of \s. \t &
\Gamma \sigmadash \Delta_2 : \s}
\\[4mm]
\infer[({\rightarrow^{\sf r}} I)]
{\Gamma \sigmadash \lambdar x \of \s.\Delta : \s\rightarrow^{\sf r} \t}
{\Gamma,x \of \s \sigmadash \Delta : \t &
\essence{\Delta} \mathrel{=_\eta} x
}
&
\infer[(\Pir E)]
{\Gamma \sigmadash \Delta_1\atr \Delta_2 : \t}
{\Gamma \sigmadash \Delta_1 : \s\rightarrow^{\sf r} \t &
\Gamma \sigmadash \Delta_2 : \s}
\\[4mm]
\multicolumn{2}{c}
{\infer[(\cap I)]
{\Gamma \sigmadash \spair{\Delta_1}{\Delta_2} : \s \cap \t}
{\Gamma \sigmadash \Delta_1 : \s &
\Gamma \sigmadash \Delta_2 : \t &
\essence{\Delta_1} \mathrel{=_\eta} \essence{\Delta_2}
}
\quad
\infer[(\cap E_l)]
{\Gamma \sigmadash \prl{\Delta} : \s}
{\Gamma \sigmadash \Delta : \s \cap \t}
\quad
\infer[(\cap E_r)]
{\Gamma \sigmadash \prr{\Delta} : \t}
{\Gamma \sigmadash \Delta : \s \cap \t}
}
\\[4mm]
\infer[(\cup I_l)]
{\Gamma \sigmadash \inll{\t} {\Delta} : \s \cup \t}
{\Gamma \sigmadash \Delta : \s
&
\Gamma \sigmadash \s \cup \t : \Type}
&
\infer[(\cup I_r)]
{\Gamma \sigmadash \inrr{\s} {\Delta} : \s \cup \t}
{\Gamma \sigmadash \Delta : \t
&
\Gamma \sigmadash \s \cup \t : \Type}
\\[4mm]
\multicolumn{2}{c}
{\infer[(\cup E)]
 {\Gamma \sigmadash
 \ssum{\Delta_1}{\Delta_2} : \Pi x \of \s\cup \t.\r}
 {\begin{array}{l@{\quad}r}
 \Gamma \sigmadash \Delta_1 : \Pi y \of \s. \r[\inll{\t}{y}/x] &
 \essence{\Delta_1} \mathrel{ =_\eta } \essence{\Delta_2}
 \\[1mm]
 \Gamma \sigmadash \Delta_2 : \Pi y \of \t. \r[\inrr{\s}{y}/x] &
 \Gamma,x\of \s\cup\t \sigmadash \r : \Type
 \end{array}}}
 \quad
 \quad
 \infer[(Conv)]
{\Gamma \sigmadash \Delta : \t}
{\begin{array}{ll}
\Gamma \sigmadash \Delta : \s & \\
\Gamma \sigmadash \t : \Type & \s =_{\Delta} \t
\end{array}
 }

\end{array}
\end{math}
\end{center}
\caption{The type rules for valid objects}
\label{OBJ}
\end{figure}
The set of rules for object formation is defined in Figure \ref{OBJ}, while the sets of rules for signatures, contexts, kinds and families are defined as usual in the Appendix: all typing rules are syntax-directed. Note that proof-functionality is enforced by the essence side-conditions in rules $({\rightarrow^{\sf r}} I)$, $(\cap I)$, and $(\cup E)$.
In the rule $(Conv)$ we rely on the external notion of equality $=_{\Delta}$.
An option could have be to add an internal notion of equality directly in the type system ($\Gamma \sigmadash \s =_{\Delta} \t$), and prove that the external and the internal definitions of equality are equivalent, as was proved for semi-full Pure Type Systems \cite{SH10}.
Yet another possibility could be to compare type essences $\essence{\s} =_{\Delta} \essence{\t}$, for a suitable extension of essence to types and kinds. Unfortunately, this would lead to undecidability of type checking, in connection with relevant implication, as the following example shows. Let the two constants $c_1$ of type $\s\rightarrow^{\sf r}(\Pi y\of\s.\s)$ and $c_2$ of type $(\Pi y\of\s.\s)\rightarrow^{\sf r}\s$: the following $\Delta$-term is typable with $\sigma$ and its essence is $\Omega$.
$$
 \Delta_\Omega \eqdef (\lambda x\of\s. c_1\atr x\at x)\at (c_2\atr (\lambda x\of\s. c_1\atr x\at x))
 \qquad \qquad
 \essence{\Delta_\Omega} = \Omega
 $$
Since the intended meaning of relevant implication is ``essentially'' the identity, introducing variables or constants whose type is a relevant implication, amounts to assuming axioms corresponding to type inclusions such as those that equate $\s$ and $\s \to \s$. As a consequence, $\beta$-equality of essences becomes undecidable. Thus, we rule out such options in relating relevant implications in \DLF\ to subtypes in the type assignment system $\B$ of \cite{BDdL}.


\subsection{Relating \DLF\ to $\B$}
We compare and contrast certain design decisions of \DLF\ to the type assignment system $\B$ of \cite{BDdL}.
The proof of strong normalization for \DLF\ will rely, in fact, on a forgetful mapping from \DLF\ to $\B$.
As pointed out in \cite{BDdL}, the elimination rule for union types in $\B$ breaks subject reduction for one-step $\beta$-reduction, but this can be recovered using a suitable parallel $\beta$-reduction. The well-known counter-example for one-step reduction, due to Pierce is
%
$$x
\at (({\sf I} \at y) \at z) \at (({\sf I} \at y) \at z) \red_\beta
\begin{array}{l}
 \nnearrow^\beta x \at (y \at z) \at (({\sf I} \at y) \at z) \ssearrow_\beta
 \\[2mm]
 \ssearrow_\beta x \at (({\sf I} \at y) \at z) \at (y \at z) \nnearrow^\beta
\end{array}
x \at (y \at z) \at (y \at z),
$$
where $\mathsf I$ is the identity.
In the typing context $B \eqdef x \of (\s_1\to\s_1\to\t) \cap (\s_2\to\s_2\to\t),y \of \r \to (\s_1\cup \s_2),z \of \r$, the first and the last terms can be typed with $\t$, while the terms in the fork cannot. The reason is that the subject in the conclusion of the $(\cup E)$ rule uses a context which can have more than one hole, as in the present case\footnote{The problem would not arise if $(\cup E)$ is replaced by the rule schema
$$ \infer[(\cup E')]
{B \vdash M[N_1/x_1 \ldots N_n/x_n]: \r}
{
B,x_1\of\s,\ldots,x_n\of\s \vdash M : \r & 
B,x_1\of\t,\ldots,x_n\of \t \vdash M : \r &
B \vdash N_i: \sigma \cup \tau \quad N_i=_\beta N_j \quad i,j=1\ldots n
}
$$
Removing the non-static clause on the $N_i$'s would yield a more permissive type system than $\B$.}.
In \DLF, the formulation of the $(\cup E)$ rule takes a different route which does not trigger the counterexample. Indeed, we have introduction and elimination constructs $\inl,\inr$ and $\ssum{ }{ }$ which allow to reduce the term only if we know that the argument, stripped of the introduction construct, has one of the types of the disjunction. Pierce's critical term can be expressed and typed in \DLF\
with the following judgment (the full derivation is in the Appendix):
$$
\Gamma
\sigmadash
\ssum{\underbrace{(\lambda x_1\of\s_1.(\prl x) \at
 x_1 \at x_1)}_{\Delta_1}}
 {\underbrace{(\lambda
 x_2\of\s_2.(\prr x) \at x_2 \at x_2)}_{\Delta_2}}
\at (\, \underbrace{(\lambda x_3\of\rho \rightarrow \s_1 \cup \s_2.x_3)}_{\Delta_3}
\at y \at z) : a\at (y \at z) \at (y \at z)$$
where $\Gamma \eqdef x \of \Pi x_1\of\s_1. \Pi x_2\of \s_1. a \at (\inll{\s_2} \at x_1) \at (\inll{\s_2} \at x_2) \cap \Pi x_1\of\s_2.\Pi x_2\of \s_2. a \at (\inrr{\s_1} \at x_1) \at (\inrr{\s_1} \at x_2),$\\ $y \of \rho \rightarrow \s_1 \cup \s_2,z\of \rho$, and $\Sigma \eqdef a\of \s_1 \cup \s_2 \rightarrow \s_1 \cup \s_2 \rightarrow \Type$.
Notice that there is only one redex, namely $\Delta_3 \at y$, and the reduction of this redex leads to $[\Delta_1, \Delta_2] \at (y\at z)$, and no other intermediate (untypable) $\Delta$-terms are possible.

The following result will be useful in the following section.
\begin{theorem}\label{th:BBDLstrongnorm}
The system $\B$ without $\omega$ gives types only to strongly normalizing terms.
\end{theorem}
A proof is embedded in Theorem~4.8 of~\cite{BDdL}. It can also be obtained using the general computability method presented in
\cite{PS} Section 4, by interpreting intersection and union types precisely as intersections and unions in the lattice of computability sets.

\subsection{\DLF\ metatheory}
\DLF\ can play the role of a Logical Framework only if decidable. Due to the lack of space, we list here only the main results: the complete list appears in the Appendix.
\noindent
The first important step states that if a $\Delta$-term is typable, then its type is unique up to $=_\Delta$.
\begin{theorem}[Unicity of types and kinds]
\hfill
\begin{enumerate}
\item If $\Gamma \sigmadash \Delta :\sigma$ and $\Gamma \sigmadash \Delta : \tau$, then $\sigma =_{\Delta} \tau$.
\item If $\Gamma \sigmadash \sigma : K$ and $\Gamma \sigmadash \sigma : K'$, then $K =_{\Delta} K'$.
\end{enumerate}
\end{theorem}

\noindent Strong normalization is proved as in \LF. First we encode \DLF-terms into terms of the type assignment system $\B$ such that redexes in the source language correspond to redexes in the target language and we use Theorem~\ref{th:BBDLstrongnorm}. Then, we introduce two forgetful mappings, namely $\erasetype{\cdot}$ and $\erase{\cdot}$, defined in Figure~\ref{fig:erase} of the Appendix, to erase dependencies in types and to drop proof-functional constructors in $\Delta$-terms and we conclude. Special care is needed in dealing with redexes occurring in type-dependencies, because these need to be flattened at the level of terms.

\begin{theorem}[Strong normalization]\hfill
\begin{enumerate}
\item \DLF\ is strongly normalizing, \ie,
\hfill
\begin{enumerate}
\item If $\Gamma \sigmadash K$, then $K$ is strongly normalizing.
\item If $\Gamma \sigmadash \s : K$, then $\sigma$ is strongly normalizing.
\item If $\Gamma \sigmadash \Delta : \s$, then $\Delta$ is strongly normalizing.
\end{enumerate}
\item Every strongly normalizing pure $\lambda$-term can be annotated so as to be the essence of a $\Delta$-term.
\end{enumerate}
\end{theorem}

\noindent
Local confluence and strong normalization entail confluence, so we have
\begin{theorem}[Confluence]
\DLF\ is confluent, \ie:
\begin{enumerate}
\item If $K_1 \red^*_{\Delta} K_2$ and $K_1 \red^*_{\Delta} K_3$, then $\exists K_4$ such that $K_2 \red^*_{\Delta} K_4$ and $K_3 \red^*_{\Delta} K_4$.

\item If $\s_1 \red^*_{\Delta} \s_2$ and $\s_1 \red^*_{\Delta} \s_3$, then $\exists \s_4$ such that $\s_2 \red^*_{\Delta} \s_4$ and $\s_3 \red^*_{\Delta} \s_4$.

\item If $\Delta_1 \red^*_{\Delta} \Delta_2$ and $\Delta_1 \red^*_{\Delta} \Delta_3$, then $\exists \Delta_4$ such that $\Delta_2 \red^*_{\Delta} \Delta_4$ and $\Delta_3 \red^*_{\Delta} \Delta_4$.
\end{enumerate}
\end{theorem}

\noindent Then, we have subject reduction, whose proof relies on technical lemmas about inversion and subderivation properties (see Appendix).

\begin{theorem}[Subject reduction of \DLF]\hfill
\begin{enumerate}
\item If $\Gamma \sigmadash K$ and $K \to_{\Delta} K'$, then $\Gamma \sigmadash K'$.
\item If $\Gamma \sigmadash \sigma : K$ and $\sigma \to_{\Delta} \sigma'$, then $\Gamma \sigmadash \sigma' : K$.
\item If $\Gamma \sigmadash \Delta : \sigma$ and $\Delta \to_{\Delta} \Delta'$, then $\Gamma \sigmadash \Delta' : \sigma$.
\end{enumerate}
\end{theorem}

Finally, we define a possible algorithm for checking judgements in \DLF\ by computing a type or a kind for a term, and then testing for definitional equality, \ie\ $=_{\Delta}$, against the given type or kind. This is achieved by reducing both to their unique normal forms and checking that they are identical up to $\alpha$-conversion. Therefore we finally have:

\begin{theorem}[Decidability]
All the type judgments of \DLF\ are recursively decidable.
\end{theorem}

\smallskip\noindent{\bf Minimal Relevant Implications and Type Inclusion.}
{Type inclusion} and the rules of subtyping are related to the notion of minimal relevant implication, see \cite{BM94,APLAS16}.
The insight is quite subtle, but ultimately very simple. This is what makes it appealing. The apparently intricate rules of subtyping and type inclusion, which occur in many systems, and might even appear \adhoc\ at times, can all be explained away in our principled approach, by proving that the relevant implication type is inhabited by a term whose essence is essentially a variable.

In the following theorem we show how relevant implication subsumes the type-inclusion rules of the theory $\Xi$ of \cite{BDdL}, without rules (5) and (13) (dealing with $\omega$) and rule (10) (distributing $\cap$ over $\cup$) in Figure \ref{curryandchurch}: we call $\Xi'$ such restricted subtype theory. Note that the reason to drop subtype rule (10) is due to the fact that we cannot inhabit the type
$\sigma \cap (\tau \cup \rho) \to^{\sf r} (\sigma \cap \tau) \cup (\sigma \cap \rho)$\footnote{To encompass also the subtype rule (10) of the type theory $\Xi$, besides adding a special constant, we can strengthen the form of the $(\cup E)$ type rule as follows:\vspace{-1ex}
\begin{scriptsize}
$$\infer[(\cup E)]
 {\Gamma \sigmadash
 \ssum{\Delta_1}{\Delta_2} : \Pi x \of \chi \cap(\s\cup \t).\r \at x}
 {\begin{array}{ll}
 \Gamma \sigmadash \Delta_1 : \Pi y \of \chi \cap\s. \r \at \spair{\prl{y}}{\inll{\t} \at \prr{y}} &
  \essence{\Delta_1} =_\eta \essence{\Delta_2}
 \\
  \Gamma \sigmadash \Delta_2 : \Pi y \of \chi \cap \t. \r \at \spair{\prl{y}}{\inrr{\s} \at \prr{y}} 
&
 \Gamma \sigmadash \r : \Pi y\of \chi \cap(\s \cup \t). \Type
 \end{array}}$$
\end{scriptsize}\vspace{-1ex}
Similarly we can treat the remaining rules of the type theory $\mathit \Pi$ in \cite{BDdL}.}.

%
%
\begin{theorem}[Type Inclusion]
The judgement $\langle \rangle\sigmadash \Delta : \s\rightarrow^{\sf r} \t $ (where both $\s$ and $\t$ do not contain dependencies or relevant families) holds iff $\s \leq \t$ holds in the subtype theory $\Xi'$ of $\B$ enriched with new axioms of the form $\s_1\leq\s_2$ for each constant $c:\s_1\rightarrow^{\sf r}\s_2\in \Sigma$.
\end{theorem}
As far as the $\lambda^{\Pi\&}$ system of Refinement Types introduced by Pfenning in~\cite{Refine93}, we have the following theorem:
\begin{corollary}[Pfenning's Refinement Types]\label{th:RefTyp}
The judgment $\vdash_\Sigma \sigma\leq\tau$ in $\lambda^{\Pi\&}$ can be encoded in \DLF\ by adding a constant of type $\sigma\rightarrow^{\sf r}\tau$ to $\Sigma'$, where the latter is the signature obtained from $\Sigma$ by replacing each clause of the form $a_1::a_2$ or $a_1\leq a_2$ in $\Sigma$ by a constant of type $a_1\rightarrow^{\sf r} a_2$.
\end{corollary}
Moreover, while Pfenning needs to add explicitly the rules of subtyping (\ie\ the theory of $\leq$) in $\lambda^{\Pi\&}$, we inherit them naturally in \DLF\ from the rules for minimal relevant implication.


\section{Examples}\label{sec:examples}

\begin{figure*}[t]
{\small
\begin{displaymath}
 \begin{array}{rcl}
 \multicolumn{3}{l}{
 \mbox{Atomic propositions, non-atomic goals and non-atomic programs:}
 \quad
 \alpha,\gamma_{0},\pi_{0} : \Type} \\

 \multicolumn{3}{l}{ \mbox{Goals and programs:}\quad
 \gamma = \alpha \cup \gamma_{0} \qquad \pi = \alpha \cup \pi_{0}} \\

 \multicolumn{3}{l}{
 \mbox{Constructors (implication, conjunction, disjunction).}} \\
 \mathsf{impl} & : & (\pi \rightarrow \gamma \rightarrow \gamma_{0}) \cap (\gamma \rightarrow \pi \rightarrow \pi_{0}) \\
 \mathsf{impl}_1 & = & \lambda x \of \pi. \lambda y \of \gamma. \inrr{\alpha}(\prl \mathsf{impl}\ x\ y)
 \qquad \mathsf{impl}_2 = \lambda x \of \gamma. \lambda y \of \pi. \inrr{\alpha}(\prr \mathsf{impl}\ x\ y) \\
 \mathsf{and} & : & (\gamma \rightarrow \gamma \rightarrow \gamma_{0}) \cap (\pi \rightarrow \pi \rightarrow \pi_{0})\\
 \mathsf{and}_1 & =& \lambda x \of \gamma. \lambda y \of \gamma. \inrr{\alpha} (\prl \mathsf{and}\ x\ y)
 \qquad\,\, \mathsf{and}_2 = \lambda x \of \pi. \lambda y \of \pi. \inrr{\alpha} (\prr \mathsf{and}\ x\ y)\\
 \mathsf{or} & : & (\gamma \rightarrow \gamma \rightarrow \gamma_{0})
 \qquad \mathsf{or}_1 = \lambda x \of \gamma. \lambda y \of \gamma. \inrr{\alpha} (\mathsf{or}\ x\ y) \\

 \multicolumn{3}{l}{
 \mbox{$\mathsf{solve}\ p\ g$ indicates that the judgment $p \vdash g$ is valid.}}\\

 \multicolumn{3}{l}{
\mbox{$\mathsf{bchain}\ p\ a\ g$ indicates that, if $p \vdash g$ is valid, then $p \vdash a$ is valid.}}\\
 \mathsf{solve} & : & \pi \rightarrow \gamma \rightarrow \Type
 \qquad
 \mathsf{bchain} : \pi \rightarrow \alpha \rightarrow \gamma \rightarrow \Type \\

 \multicolumn{3}{l}{
 \mbox{Rules for $\mathsf{solve}$:}} \\
 - & : & \Pi_{(p \of \pi) (g_{1}, g_{2} \of \gamma)} \mathsf{solve}\ p\ g_{1} \rightarrow \mathsf{solve}\ p\ g_{2} \rightarrow \mathsf{solve}\ p\ (\mathsf{and}_1\ g_{1}\ g_{2}) \\
 - & : & \Pi_{(p \of \pi) (g_{1}, g_{2} \of \gamma)} \mathsf{solve}\ p\ g_{1} \rightarrow \mathsf{solve}\ p\ (\mathsf{or}_1\ g_{1}\ g_{2}) \\
 - & : & \Pi_{(p \of \pi) (g_{1}, g_{2} \of \gamma)} \mathsf{solve}\ p\ g_{2} \rightarrow \mathsf{solve}\ p\ (\mathsf{or}_1\ g_{1}\ g_{2}) \\
 - & : & \Pi_{(p_{1}, p_{2} \of \pi) (g \of \gamma)} \mathsf{solve}\ (\mathsf{and}_2\ p_{1}\ p_{2})\ g \rightarrow \mathsf{solve}\ p_{1}\ (\mathsf{impl}_1\ p_{2}\ g) \\
 - & : & \Pi_{(p \of \pi)(a \of \alpha)(g \of \gamma)} \mathsf{bchain}\ p\ a\ g \rightarrow \mathsf{solve}\ p\ g \rightarrow \mathsf{solve}\ p\ (\inll{\gamma_0} a) \\

 \multicolumn{3}{l}{
 \mbox{Rules for $\mathsf{bchain}$:}} \\
 - & : & \Pi_{(a \of \alpha)(g \of \gamma)} \mathsf{bchain}\ (\mathsf{impl}_2\ g\ (\inl{\pi_0} a))\ a\ g \\
 - & : & \Pi_{(p_{1}, p_{2} \of \pi)(a \of \alpha)(g \of \gamma)} \mathsf{bchain}\ p_{1}\ a\ g \rightarrow \mathsf{bchain}\ (\mathsf{and}_2\ p_{1}\ p_{2})\ a\ g \\
 - & : & \Pi_{(p_{1}, p_{2} \of \pi)(a \of \alpha)(g \of \gamma)} \mathsf{bchain}\ p_{2}\ a\ g \rightarrow \mathsf{bchain}\ (\mathsf{and}_2\ p_{1}\ p_{2})\ a\ g \\
 - & : & \Pi_{(p \of \pi)(a \of \alpha)(g, g_{1}, g_{2} \of \gamma)} \mathsf{bchain}\ (\mathsf{impl}_2\ (\mathsf{and}_1\ g_{1}\ g_{2})\ p)\ a\ g \rightarrow \mathsf{bchain}\ (\mathsf{impl}_2\ g_{1}\ (\mathsf{impl}_2\ g_{2}\ p))\ a\ g \\
 - & : & \Pi_{(p_{1}, p_{2} \of \pi)(a \of \alpha)(g, g_{1} \of \gamma)} \mathsf{bchain}\ (\mathsf{impl}_2\ g_{1}\ p_{1})\ a\ g \rightarrow \mathsf{bchain}\ (\mathsf{impl}_2\ g_{1}\ (\mathsf{and}_2\ p_{1}\ p_{2}))\ a\ g \\
 - & : & \Pi_{(p_{1}, p_{2} \of \pi)(a \of \alpha)(g, g_{1} \of \gamma)}, \mathsf{bchain}\ (\mathsf{impl}_2\ g_{1}\ p_{2})\ a\ g \rightarrow \mathsf{bchain}\ (\mathsf{impl}_2\ g_{1}\ (\mathsf{and}_2\ p_{1}\ p_{2}))\ a\ g
 \end{array}
 \end{displaymath}
 }
 \caption{The \DLF\ encoding of Hereditary Harrop Formul\ae}
 \label{HHF}
 \end{figure*}

\noindent As we have argued in the previous sections, the point of this paper is a uniform and principled approach to the encoding of a plethora of type disciplines and systems which ultimately stem or can capitalize from strong proof-functional connectives and subtyping. The framework \DLF, presented in this paper, is the first to accommodate all the examples and counterexamples that have appeared in the literature. The complete developments of both the implementation of the $\Delta$-framework and example encodings can be found in \cite{BULL}.

We start the section showing the expressive power of \DLF\ in encoding classical features of typing disciplines with strong intersection and union.

\noindent {\bf Auto application.} The judgement $\vdash_{\B}\lambda x. x \at x :\s \cap (\s \to \t) \to \t$ in $\B$, is rendered in \DLF\ by the \DLF-judgement $ \sigmadash \lambda x \of\s \cap (\s \to \t). (\prr x) \at (\prl x): \s \cap (\s \to \t) \to \t$.

\noindent {\bf Polymorphic identity.} The judgement $\vdash_{\B}\lambda x.x:(\s \to \s) \cap (\t \to \t)$ in $\B$, is rendered in \DLF\ by the judgement
$\vdash_{\langle\rangle}
\spair{\lambda x \of \s.x}{\lambda x\of \t.x}: (\s \to \s) \cap (\t \to \t)
$.

\noindent {\bf Commutativity of union.} The judgement $\lambda x.x :(\s \cup \t) \to (\t \cup \s)$ in $\B$ is rendered in \DLF\ by the judgement
$
\lambda x\of\s{\cup}\t. \ssum{\lambda y\of\s. \inrr{\t} y}{\lambda y\of\t. \inll{\s} \at y} \at x :(\s \cup \t) \to (\t \cup \s).
$

\noindent \textbf{\bf Pierce's expression of page 2.} The expressive power of union types highlighted by Pierce is rendered in \DLF\ by\vspace{-1.5ex}
$$
\begin{array}{rcl}
\multicolumn{3}{l}
{Neg : \Type \qquad Zero : \Type \qquad Pos : \Type \qquad
 T : \Type \qquad F : \Type \qquad
 \mbox{Test} : Pos \cup Neg} \\

 \mbox{Is}\_0 & : & (Neg \rightarrow F) \cap (Zero \rightarrow T) \cap (Pos \rightarrow F) \\

 \mbox{Is}\_0\_\mbox{Test} & \eqdef &
 \ssum{\lambda x \of Neg. (\prl{\prl{\mbox{Is}\_0}}) \at x}{\lambda x \of Pos. (\prr{\mbox{Is}\_0}) \at x} \at \mbox{Test}
\end{array}
$$
The above example illustrates the advantages of taking \DLF\ as a framework. In \LF\ we would render it only encoding $\B$ deeply, ending up with the verbose code in {\color{blue} \href{https://github.com/cstolze/Bull/blob/master/coq_encodings/pierce_program.v}{pierce\_program.v}} \cite{BULL}.

\smallskip \noindent \textbf{Hereditary Harrop Formul\ae.}
The encoding of Hereditary Harrop's Formul\ae\ is one of the motivating examples given by Pfenning for introducing refinement types in \cite{Refine93}. In \DLF\ it can be expressed as in Figure \ref{HHF}  and type checked in \cite{LFMTP17} using our concrete syntax (file {\color{blue} \href{https://github.com/cstolze/Bull/blob/master/bull/pfenning_harrop.bull}{pfenning\_harrop.bull}} \cite{BULL}), without any reference to intersection types, by a subtle use of union types. We add also rules for solving and backchaining. Hereditary Harrop formul\ae\ can be recursively defined using two mutually recursive syntactical objects called programs ($\pi$) and goals ($\gamma$):\vspace{-1.5ex}
$$\begin{array}{rcl}
 \gamma & := & \alpha \mid \gamma \wedge \gamma \mid \pi \Rightarrow \gamma \mid \gamma \vee \gamma \\
 \end{array}
 \qquad
 \begin{array}{rcl}
 \pi & := & \alpha \mid \pi \wedge \pi \mid \gamma \Rightarrow \pi
 \end{array}
$$
Using Theorem~\ref{th:RefTyp}, we can provide an alternative encoding of atoms, goals and programs which is more faithful to the one by Pfenning. Namely, we can introduce in the signature the constants $c_1:\alpha\rightarrow^{\sf r}\gamma$ and $c_2:\alpha\rightarrow^{\sf r}\pi$ in order to represent the axioms $atom\leq goal$ and $atom\leq prog$ in Pfenning's encoding. Our approach based on union types, while retaining the same expressivity permits to shortcut certain inclusions and to rule out also certain exotic goals and exotic programs. Indeed, for the purpose of establishing the adequacy of the encoding, it is sufficient to avoid variables involving union types in the derivation contexts.


\smallskip \noindent \textbf{Natural Deductions in Normal Form.}
The second motivating example for intersection types given in~\cite{Refine93} is {\em natural deductions in normal form}. We recall that a natural deduction is in normal form if there are no applications of elimination rules of a logical connective immediately following their corresponding introduction, in the main branch of a subderivation.

\noindent The encoding we give in \DLF\ is a slightly improved version of the one in~\cite{Refine93}: as Pfenning, we restrict to the purely implicational fragment. As in the previous example, we use
\begin{wrapfigure}{l}{0.62\textwidth}
\vspace{-3mm}
$\begin{array}{rcl}
o &:& Type\quad
\supset : o\to o\to o\quad
Elim, {\rm Nf}^0 : o\to\Type\quad\\
{\rm Nf} & \equiv & \Pi A\of o.{\rm Nf}^0(A)\cup Elim(A)\\
\supset_I & : & \Pi A,B\of o.(Elim(A)\to {\rm Nf}(B))\to {\rm Nf}^0(A\supset B)\\
\supset_E & : & \Pi A,B\of o.Elim(A\supset B)\to {\rm Nf}^0(A)\to Elim(B).
\end{array}$
\vspace{-3mm}
\end{wrapfigure}
union types to define normal forms (${\rm Nf}(A)$) either as pure elimination-deductions from hypotheses ($Elim(A)$) or normal form-deductions (${\rm Nf}^0(A)$). As above we could have used also intersection types. This example is interesting in itself, being the prototype of the encoding of type systems using canonical and atomic syntactic categories \cite{harper-licata} and also of Fitch Set Theory \cite{Fitch-APLAS16}.


\lstset{backgroundcolor={\color{white}}}
\begin{figure}[t]
\begin{coq}
(* Define our types *)
Axiom o : Set.
(* Axiom omegatype : o. *)
Axioms (arrow inter union : o -> o -> o).

(* Transform our types into LF types *)
Axiom OK : o -> Set.

(* Define the essence equality as an equivalence relation *)
Axiom Eq : forall (s t : o), OK s -> OK t -> Prop.
Axiom Eqrefl : forall (s : o) (M : OK s), Eq s s M M.
Axiom Eqsymm : forall (s t : o) (M : OK s) (N : OK t), Eq s t M N -> Eq t s N M.
Axiom Eqtrans : forall (s t u : o) (M : OK s) (N : OK t) (O : OK u), Eq s t M N -> Eq t u N O -> Eq s u M O.

(* constructors for arrow (->I and ->E) *)
Axiom Abst : forall (s t : o), ((OK s) -> (OK t)) -> OK (arrow s t).
Axiom App : forall (s t : o), OK (arrow s t) -> OK s -> OK t.

(* constructors for intersection *)
Axiom Proj_l : forall (s t : o), OK (inter s t) -> OK s.
Axiom Proj_r : forall (s t : o), OK (inter s t) -> OK t.
Axiom Pair : forall (s t : o) (M : OK s) (N : OK t), Eq s t M N -> OK (inter s t).

(* constructors for union *)
Axiom Inj_l : forall (s t : o), OK s -> OK (union s t).
Axiom Inj_r : forall (s t : o), OK t -> OK (union s t).
Axiom Copair : forall (s t u : o) (X : OK (arrow s u)) (Y : OK (arrow t u)), OK (union s t) ->
 Eq (arrow s u) (arrow t u) X Y -> OK u.

(* define equality wrt arrow constructors *)
Axiom Eqabst : forall (s t s' t' : o) (M : OK s -> OK t) (N : OK s' -> OK t'),
 (forall (x : OK s) (y : OK s'), Eq s s' x y -> Eq t t' (M x) (N y)) ->
 Eq (arrow s t) (arrow s' t') (Abst s t M) (Abst s' t' N).
Axiom Eqapp : forall (s t s' t' : o) (M : OK (arrow s t)) (N : OK s) (M' : OK (arrow s' t')) (N' : OK s'),
 Eq (arrow s t) (arrow s' t') M M' -> Eq s s' N N' -> Eq t t' (App s t M N) (App s' t' M' N').

(* define equality wrt intersection constructors *)
Axiom Eqpair : forall (s t : o) (M : OK s) (N : OK t) (pf : Eq s t M N), Eq (inter s t) s (Pair s t M N pf) M.
Axiom Eqproj_l : forall (s t : o) (M : OK (inter s t)), Eq (inter s t) s M (Proj_l s t M).
Axiom Eqproj_r : forall (s t : o) (M : OK (inter s t)), Eq (inter s t) t M (Proj_r s t M).

(* define equality wrt union *)
Axiom Eqinj_l : forall (s t : o) (M : OK s), Eq (union s t) s (Inj_l s t M) M.
Axiom Eqinj_r : forall (s t : o) (M : OK t), Eq (union s t) t (Inj_r s t M) M.
Axiom Eqcopair : forall (s t u : o) (M : OK (arrow s u)) (N : OK (arrow t u)) (O : OK (union s t))
 (pf: Eq (arrow s u) (arrow t u) M N) (x : OK s),
 Eq s (union s t) x O -> Eq u u (App s u M x) (Copair s t u M N O pf).
\end{coq}\vspace{-2.5ex}
\caption{The LF encoding of $\B$ (Coq syntax)}
\label{BDDLINCOQ}
\end{figure}\vspace{-2ex}

\smallskip \noindent \textbf{Adequacy, Canonical Forms, Exotic terms.}
In the presence of union types, we have to pay special attention to the exact formulation of Adequacy Theorems, as in the Harrop's formul\ae\ example above. Otherwise exotic terms arise, such as $\ssum{\lambda x\of \sigma.C(x)}{\lambda x\of\tau.D(x)}\at y$, where $C(\cdot)$ and $D(\cdot)$ are distinct contexts (\ie\ terms with holes), which cannot be naturally simplified even if $\essence{C}\equiv \essence{D}$. More work needs to be done to streamline how to exclude, or even capitalize on exotic terms.\vspace{-.5ex}

\smallskip \noindent \textbf{Metacircular Encodings.} The following diagram summarizes the network of adequate encodings/inclusions between \DLF, \LF, and $\B$ that can be defined. We denote by ${\cal S}_1 \Longrightarrow {\cal S}_2$
\begin{wrapfigure}{l}{0.3\textwidth}
\vspace{-2mm}
\centering
$\xymatrix{
 \mathrm{LF}\ar@{=>}[r]^{sh} & \mathrm{LF}_\Delta\ar@{=>}[r]^{dp} & \mathrm{LF}\\
 \mathcal{B}\ar@{=>}[ru]^{sh}\ar@{=>}[r]^{dp} & \mathrm{LF}\ar@{^{(}->}[u] &
}$
\vspace{-6mm}
\end{wrapfigure}
%
%
%
\noindent
the encoding of system ${\cal S}_1$ in system ${\cal S}_2$, where the label \textit{sh} (resp. \textit{dp}), denotes a shallow (resp. deep) embedding. The notation ${\cal S}_1\hookrightarrow {\cal S}_2$ denotes that ${\cal S}_2$ is an extension of ${\cal S}_1$.
Due to lack of space, but with the intention of providing a better formal understanding of the semantics of strong intersection and union types in a logical framework, we provide in Figure \ref{BDDLINCOQ} a deep \LF\ encoding of a presentation of $\B$ \ala\ Church \cite{APLAS16}.
A shallow encoding of $\B$ in \DLF\ (file \BLUE{\href{https://github.com/cstolze/Bull/blob/master/bull/intersection_union.bull}{intersection\_union.bull}} \cite{BULL}) can be mechanically type checked in \cite{LFMTP17}.
A shallow encoding of \LF\ in \DLF\ (file \BLUE{\href{https://github.com/cstolze/Bull/blob/master/bull/lf.bull}{lf.bull}}) making essential use of intersection types can be also type checked.




%

\smallskip \noindent \textbf{LF encoding of $\B$.}
Figure \ref{BDDLINCOQ} presents a pure LF encoding of a presentation of $\B$ \ala\ Church in Coq syntax using HOAS. We use HOAS in order to take advantage of the higher-order features of the frameworks: other abstract syntax representation techniques would not be much different, but more verbose.
The \texttt{Eq} predicate plays the same role of the essence function in \DLF, namely, it encodes the judgement that two proofs (\ie\ two terms of type \texttt{(OK \_)}) have the same structure. This is crucial in the \texttt{Pair} axiom (\ie\ the introduction rule of the intersection type constructor) where we can inhabit the type \texttt{(inter s t)} only when the proofs of its component types \texttt{s} and \texttt{t} share the same structure (\ie\ we have a witness of type \texttt{(Eq s t M N)}, where \texttt{M} has type \texttt{(OK s)} and \texttt{N} has type \texttt{(OK t)}). A similar role is played by the \texttt{Eq} premise in the \texttt{Copair} axiom (\ie\ the elimination rule of the union type constructor). We have an \texttt{Eq} axiom for each proof rule. Examples of this encoding can be found in {\color{blue} \href{https://github.com/cstolze/Bull/blob/master/coq_encodings/intersection_union.v}{intersection\_union.v}} \cite{BULL}.
\vspace{-1.5ex}



\section{Implementation and Future Work}\label{sec:impl}
In a previous paper \cite{LFMTP17}, we have implemented in OCaml suitable algorithms for type reconstruction, as well as type checking. In \cite{TTCS17} we have implemented the subtyping algorithm which extends the well-known Hindley algorithm for intersection types \cite{Hindley82} with union types. The subtyping algorithm has been mechanically proved correct in Coq, extending the Bessai's mechanized proof of a subtyping algorithm for intersection types \cite{bessai16}.


A Read-Eval-Print-Loop allows to define axioms and definitions, and performs some basic terminal-style features like error pretty-printing, subexpressions highlighting, and file loading. Moreover, it can type-check a proof or normalize it, using a strong reduction evaluator. We use the syntax of Pure Type Systems \cite{berardi} to improve the compactness and the modularity of the kernel. Binders are implemented using de Brujin indexes. We implemented the conversion rule in the simplest way possible: when we need to compare types, we syntactically compare their normal form. Abstract and concrete syntax are mostly aligned: the concrete syntax is similar to the concrete syntax of Coq
(see {\color{blue} \href{https://github.com/cstolze/Bull}{Bull}} and {\color{blue} \href{https://github.com/cstolze/Bull-Subtyping}{Bull-Subtyping}} \cite{BULL}).

We are currently designing a higher-order unification algorithm for $\Delta$-terms and a bidirectional refinement algorithm, similar to the one found in \cite{Refine12}. The refinement can be split into two parts: the essence refinement and the typing refinement. In the same way, there will be a unification algorithm for the essence terms, and a unification algorithm for $\Delta$-terms.
The bidirectional refinement algorithm aims to have partial type inference, and to give as much information as possible to a hypothetical solver, or the unifier. For instance, if we want to find a $?y$ such that $\sigmadash \spair{\lambda x\of\s.x}{\lambda x\of\t.?y} : (\s \rightarrow \s) \cap (\t \rightarrow \t)$, we can infer that $x\of\t\vdash ?y : \t$ and that $\essence{?y} = x$.



\bibliographystyle{plainurl}
\bibliography{inter_union_biblio_1.0}

\appendix


\section{Appendix} 
Let Figure \ref{SIGCTX} denote Valid Signatures and Contexts and Figure \ref{KFAM} denote Valid Kinds and Families.

\begin{figure}[h!]
\begin{center}
\begin{math}
\begin{array}{r@{\qquad}r}
\multicolumn{2}{l}
{\mbox{Let $\Gamma\eqdef
\{ x _1\of \s_1,\ldots, x _n\of \s_n\}~(i \neq j$ implies
$
x _i \not \equiv x _j)$, and $\Gamma, x \of\s \eqdef \Gamma \cup \{ x \of\s\}$}}
\\[2mm]
\multicolumn{2}{l}
{\mbox{Let $\Sigma\eqdef
 \{ c_1\of \s_1,\ldots, c_n\of \s_n\}$,
 and
$\Sigma, c \of\s \eqdef \Sigma \cup \{ c \of\s\}$}}
\\[2mm]
\multicolumn{2}{l}{\mbox{Valid Signatures}}
\\[2mm]
\multicolumn{2}{c}
{\infer[(\epsilon \Sigma)]
{\oksig \emptysig}
{}
\qquad
\infer[(K \Sigma)]{\oksig {\Sigma, a \of K}}
{\oksig {\Sigma} & \sigmadash K & a \not \in \dom \Sigma}
\qquad
\infer[(\s \Sigma)]
{\oksig {\Sigma, c \of \s}}
{\oksig {\Sigma} & \sigmadash \s : \Type & c \not \in \dom \Sigma}
}\\
\\[4mm]
\multicolumn{2}{l}{\mbox{Valid Contexts}}
\\[2mm]
\infer[(\epsilon \Gamma)]
{\sigmadash \emptyctx}
{\oksig \Sigma}
&
\infer[(\s \Gamma)]
{\sigmadash \Gamma, x \of \s}
{\sigmadash \Gamma & \Gamma \sigmadash \s : \Type & x \not \in \dom \Gamma}
\end{array}
\end{math}
\end{center}
\caption{Valid Signatures and Contexts}
\label{SIGCTX}
\end{figure}

\begin{figure}[h!]
\begin{center}
\begin{math}
\begin{array}{r@{\qquad}r}
\multicolumn{2}{l}{\mbox{Valid Kinds}}
\\[2mm]
\infer[(Type)]
{\Gamma \sigmadash \Type}
{\sigmadash \Gamma}
&
\infer[(\Pi K)]
{\Gamma \sigmadash \Pi x \of \s. K}
{\Gamma, x \of \s \sigmadash K}
\\[4mm]
\multicolumn{2}{l}{\mbox{Valid Families}}
\\[2mm]
\infer[(Const)]
{\Gamma \sigmadash a : K}
{\sigmadash \Gamma & a \of K \in \Sigma}
&
\infer[(Conv)]
{\Gamma \sigmadash \s : K_2}
{\Gamma \sigmadash \s : K_1 & \Gamma \sigmadash K_2 & K_1 =_{\Delta} K_2}
\\[2mm]
\infer[(\Pi I)]
{\Gamma \sigmadash \Pi x \of \s. \t : \Type}
{\Gamma, x \of \s \sigmadash \t : \Type}
&
\infer[(\Pi E)]
{\Gamma \sigmadash \s \at \Delta : K [\Delta/ x ]}
{\Gamma \sigmadash \s : \Pi x \of \t. K & \Gamma \sigmadash \Delta : \t}
\\[4mm]
\infer[(\rightarrow^{\sf r} I)]
{\Gamma \sigmadash \s\rightarrow^{\sf r} \t : \Type}
{\Gamma \sigmadash \s :\Type & \Gamma \sigmadash \t : \Type}
&
\\[4mm]
\infer[(\cap I)]
{\Gamma \sigmadash \s \cap\t : \Type}
{\Gamma \sigmadash \s : \Type & \Gamma \sigmadash \t : \Type}
&
\infer[(\cup I)]
{\Gamma \sigmadash \s \cup \t : \Type}
{\Gamma \sigmadash \s : \Type & \Gamma \sigmadash \t : \Type}
\end{array}
\end{math}
\end{center}
\caption{Valid Kinds and Families}
\label{KFAM}
\end{figure}

\DLF\ can play the role of a logical framework only if decidable. The road map which we follow to establish decidability is the standard one, see \eg\ \cite{LF}. In particular, we prove in order: uniqueness of types and kinds, structural properties, normalization for raw well-formed terms, and hence confluence. Then we prove the inversion property, the subderivation property, subject reduction, and finally decidability. 
\begin{lemma}
Let $\alpha$ be either $\s : K$ or $\Delta: \s$. Then:
\begin{enumerate}
\item Weakening: If $\Gamma \sigmadash \alpha$ and $\sigmadash \Gamma,\Gamma'$, then $\Gamma,\Gamma' \sigmadash \alpha$.
\item Strengthening: If $\Gamma,x\of\sigma,\Gamma'\sigmadash \alpha$, then $\Gamma,\Gamma' \sigmadash \alpha$, provided that $x \not \in FV(\Gamma') \cup FV(\alpha)$.
\item Transitivity: If $\Gamma \sigmadash \Delta : \sigma$ and $\Gamma,x\of\sigma,\Gamma'\sigmadash \alpha$, then $\Gamma,\Gamma'[\Delta/x] \sigmadash \alpha[\Delta/x]$.
\item Permutation: If $\Gamma,x_1\of\sigma,\Gamma',x_2\of\tau,\Gamma''\sigmadash \alpha$, then
$\Gamma,x_2\of\tau,\Gamma',x_1\of\sigma,\Gamma'' \\ \sigmadash \alpha$,
provided that $x_1$ does not occur free in $\Gamma'$ or in $\tau$, and that $\tau$ is valid in $\Gamma$.
\end{enumerate}
\end{lemma}

\addtocounter{theorem}{-8}

\begin{theorem}[Unicity of Types and Kinds]
\hfill
\begin{enumerate}
\item If $\Gamma \sigmadash \Delta :\sigma$ and $\Gamma \sigmadash \Delta : \tau$, then $\sigma =_{\Delta} \tau$.
\item If $\Gamma \sigmadash \sigma : K$ and $\Gamma \sigmadash \sigma : K'$, then $K =_{\Delta} K'$.
\end{enumerate}
\end{theorem}

\addtocounter{theorem}{7}

In order to prove strong normalization we follow the pattern used for pure \LF. Namely, we map \DLF-terms into terms of the system $\B$ in such a way that redexes in the source language are mapped into redexes in the target language, and then take advantage of Theorem~\ref{th:BBDLstrongnorm}. Special care is needed in dealing with redexes occurring in type-dependencies, because these need to be flattened at the level of terms.

\begin{definition}\label{def:forgetful}
Let the forgetful mappings $\erasetype{\cdot}$ and $\erase{\cdot}$ be defined as in Figure~\ref{fig:erase}.
\begin{figure}
$$
\begin{array}{rcl}
\erasetype{\Type} & = & \top\quad\mbox{(a special constant)}\\[1mm]
\erasetype{\Pi x\of \s.K} & = & \erasetype{\s}\to\erasetype{K}\\[1mm]
\erasetype{a} & = & a\\[1mm]
\erasetype{\Pi x\of\s.\t} & = & \erasetype{\s}\to\erasetype{\t}\\[1mm]
\end{array}
\begin{array}{rcl}
\erasetype{\s\rightarrow^{\sf r}\t} & = & \erasetype{\s}\to\erasetype{\t}\\[1mm]
\erasetype{\s\at\Delta} & = & \erasetype{\s}\\[1mm]
\erasetype{\s\cap\t} & = & \erasetype{\s}\cap\erasetype{\t}\\[1mm]
\erasetype{\s\cup\t} & = & \erasetype{\s}\cup\erasetype{\t}
\end{array}
$$

\begin{minipage}{0.5\textwidth}
\begin{eqnarray*}
\erase{a} & = & a \\[1mm]
\erase{c} & = & c\\[1mm]
\erase{x} & = & x\\[1mm]
\erase{\s\at\Delta} & = & \erase{\s}\at\erase{\Delta}\\[1mm]
\erase{\Delta_1\at\Delta_2} & = & \erase{\Delta_1}\at\erase{\Delta_2}\\[1mm]
\erase{\Delta_1\atr\Delta_2} & = & \erase{\Delta_1}\at\erase{\Delta_2}\\[1mm]
\erase{\lambda x\of\s.\Delta} & = & (\lambda y.\lambda x.\erase{\Delta})\erase{\s}\mbox{ $y\not\in fv(\Delta)$}\\[1mm]
\erase{\lambdar x\of\s.\Delta} & = & (\lambda y.\lambda x.\erase{\Delta}) \erase{\s}\mbox{ $y\not\in fv(\Delta)$}\\[1mm]
\erase{\Pi x\of\s.\t} & = & c_{\erasetype{\s}}\at\erase{\s}\at(\lambda x.\erase{\t})
\end{eqnarray*}
\end{minipage}
\quad
\begin{minipage}{0.5\textwidth}
\begin{eqnarray*}
\erase{\s\rightarrow^{\sf r}\t} & = & c_\times \at \erase{\s} \at \erase{\t}\\[1mm]
%
%
\erase{\s\cap\t} & = & c_\times \at \erase{\s} \at \erase{\t}\\[1mm]
\erase{\s\cup\t} & = & c_\times \at \erase{\s} \at \erase{\t}\\[1mm]
\erase{\spair{\Delta_1}{\Delta_2}} & = & \erase{{\Delta_1}}\\[1mm]
\erase{\ssum{\Delta_1}{\Delta_2}} & = & \erase{\Delta_1}\\[1mm]
\erase{\prl{\Delta}} & = & \erase{\Delta}\\[1mm]
\erase{\prr{\Delta}} & = & \erase{\Delta}\\[1mm]
\erase{\inll{\s}{\Delta}} & = & (\lambda x. \erase{\Delta}) \at \erase{\s}\mbox{ $x\not\in fv(\Delta)$} \\[1mm]
\erase{\inrr{\s}{\Delta}} & = & (\lambda x. \erase{\Delta}) \at \erase{\s} \mbox{ $x\not\in fv(\Delta)$}
\end{eqnarray*}
\end{minipage}
\caption{The forgetful mappings $\erasetype{\cdot}$ and $\erase{\cdot}$}\label{fig:erase}
\end{figure}
\end{definition}

The forgetful mappings are extended to contexts and signatures in the obvious way. The clauses for strong pairs/co-pairs are justified by the following lemma:
\begin{lemma}\label{compilred}
If $\Gamma\sigmadash \spair{\Delta_1}{\Delta_2} : \s$ or $\Gamma\sigmadash \ssum{\Delta_1}{\Delta_2} : \s$, then $\erase{\Delta_1}=_\beta\erase{\Delta_2}$.
\end{lemma}
The following lemmas are proved by straightforward structural induction.

\begin{lemma}\hfill
\begin{enumerate}
\item If $\s=_{\Delta}\t$, then $\erasetype{\s}=_\beta\erasetype{\t}$.
\item If $K_1=_{\Delta}K_2$, then $\erasetype{K_1}=_\beta\erasetype{K_2}$.
\end{enumerate}
\end{lemma}

\begin{lemma}\hfill
\begin{enumerate}
\item $\erase{\Delta_1[\Delta_2/x]}=_\beta \erase{\Delta_1}[\erase{\Delta_2}/x]$.
\item $\erase{\s[\Delta/x]}=_\beta \erase{\s}[\erase{\Delta}/x]$.
\end{enumerate}
\end{lemma}

\begin{lemma}\label{lemma:maptyping}\hfill
\begin{enumerate}
\item If $\Gamma\sigmadash \s : K$, then $\erasetype{\Gamma}\vdash_{\B^+} \erase{\s} \ : \ \erasetype{K}$.


\item If $\Gamma\sigmadash \Delta : \s$, then $\erasetype{\Gamma}\vdash_{\B^+} \erase{\Delta} \ : \ \erasetype{\s}$.


\end{enumerate}
where $\vdash_{\B^+}$ denotes the type system $\B$, augmented by $c_\times:\top\to \top\to \top$ and the infinite set of axioms 
$c_{\erasetype{\s}}:\top\to(\erasetype{\s}\to\top)\to\top
$,
for each type $\s$.


\end{lemma}
Notice that the function $\essence{\ }$ and $\erase{\ }$ treat differently relevant implication.

\begin{lemma}\label{lemma:mapreductions}\hfill
\begin{enumerate}
\item If $\s\red_{\beta}\t$, then $\erase{\s}\red^+_\beta\erase{\t}$.
\item if $\Delta_1\red_{\beta} \at \Delta_2$, then $\erase{\Delta_1}\red^+_\beta\erase{\Delta_2}$.
\end{enumerate}
\end{lemma}

Parallel reduction enjoys the strong normalization property, \ie\

\addtocounter{theorem}{-14}

\begin{theorem}[Strong normalization]\label{th:strongnorm} \hfill
\begin{enumerate}
\item The \DLF\ is strongly normalizing, \ie,
\hfill
\begin{enumerate}
\item If $\Gamma \sigmadash K$, then $K$ is strongly normalizing.
\item If $\Gamma \sigmadash \s : K$, then $\sigma$ is strongly normalizing.
\item If $\Gamma \sigmadash \Delta : \s$, then $\Delta$ is strongly normalizing.
\end{enumerate}
\item Every strongly normalizing pure $\lambda$-term can be annotated so as to be the essence of a $\Delta$-term.
\end{enumerate}
\end{theorem}
 
\addtocounter{theorem}{13}

\begin{proof}
1) Strong normalization derives directly from Lemmas~\ref{lemma:maptyping}, \ref{lemma:mapreductions} and Theorem~\ref{th:BBDLstrongnorm}.\\
2) By induction on the specification of strongly normalizing terms which can be inductively defined as
i) $\Delta_1 \ldots \Delta_n \in SN \Rightarrow \lambda x_1, \ldots, x_n. x\at \Delta_1 \ldots \Delta_n \in SN$ for $x$ possibly among the $x_i$'s,
 ii) $\Delta [\Delta'/x] \at \Delta_1 \ldots \Delta_n \in SN$, and
 iii) $\Delta' \in SN$
 $\Rightarrow (\lambda x\of\s.\Delta) \at \Delta' \at \Delta_1 \at \ldots \at \Delta_n \in SN$.
\end{proof}


Local confluence (Proposition~\ref{prop:localconfluence}) and strong normalization (Theorem~\ref{th:strongnorm}) entail confluence, so we have
\begin{theorem}[Confluence] \label{th:confluence}
\DLF\ is confluent, \ie:
\begin{enumerate}
\item If $K_1 \red^*_{\Delta} K_2$ and $K_1 \red^*_{\Delta} K_3$, then $\exists K_4$ such that $K_2 \red^*_{\Delta} K_4$ and $K_3 \red^*_{\Delta} K_4$.

\item If $\s_1 \red^*_{\Delta} \s_2$ and $\s_1 \red^*_{\Delta} \s_3$, then $\exists \s_4$ such that $\s_2 \red^*_{\Delta} \s_4$ and $\s_3 \red^*_{\Delta} \s_4$.

\item If $\Delta_1 \red^*_{\Delta} \Delta_2$ and $\Delta_1 \red^*_{\Delta} \Delta_3$, then $\exists \Delta_4$ such that $\Delta_2 \red^*_{\Delta} \Delta_4$ and $\Delta_3 \red^*_{\Delta} \Delta_4$.
\end{enumerate}
\end{theorem}
The following lemmas are proved by structural induction.

\begin{lemma}[Inversion properties]
\label{prp:inversion}
\hfill
\begin{enumerate}
\item If ${\Pi x\of\sigma. \t} =_{\Delta} \t''$, then $\t'' \equiv {\Pi x\of\s'.\t'}$, for some $\s',\t'$, such that $\s' =_{\Delta} \s$, and $\t' =_{\Delta} \t$.

\item If $\s\rightarrow^{\sf r} \t =_{\Delta} \t''$, then $\t'' \equiv \s'\rightarrow^{\sf r} \t'$, for some $\s',\t'$, such that $\s' =_{\Delta} \s$, and $\t' =_{\Delta} \t$.

\item If $\s\cap\t =_{\Delta}\r$, then $\r \equiv \s'\cap\t'$, for some $\s',\t'$, such that $\s' =_{\Delta} \s$, and $\t' =_{\Delta} \t$.

\item If $\s\cup\t =_{\Delta}\r$, then $\r \equiv \s'\cup\t'$, for some $\s',\t'$, such that $\s' =_{\Delta} \s$, and $\t' =_{\Delta} \t$.

\item If $\Gamma \sigmadash \lambda x\of\s.\Delta : \Pi x\of\s.\t$, then $\Gamma, x \of \s \sigmadash \Delta : \t$.

\item If $\Gamma \sigmadash \lambdar x\of\s.\Delta : \Pi x\of\s.\t$, then $\Gamma, x \of \s \sigmadash \Delta : \t$ and $\essence{\Delta}=_\eta x$.

\item If $\Gamma \sigmadash \spair{\Delta_1}{\Delta_2} : \s\cap\t$, then $\Gamma \sigmadash \Delta_1 : \s$, $\Gamma \sigmadash \Delta_2 : \t$, and $\essence{\Delta_1}=_\beta\essence{\Delta_2}$.


\item If $\Gamma \sigmadash \ssum{\Delta_1}{\Delta_2} : \Pi x\of\s \cup \t.\rho$, then $\Gamma \sigmadash \Delta_1 : \Pi y\of\s.\rho \at (\inll{\t} \at y)$, $\Gamma \sigmadash \Delta_2 : \Pi y\of\t.\rho \at (\inrr{\s} \at y)$, and $\essence{\Delta_1}=_\beta \essence{\Delta_2}$.

\item If $\Gamma \sigmadash \prl{\Delta} : \s$, then $\Gamma \sigmadash \Delta\of\s\cap\t$, for some $\t$.

\item If $\Gamma \sigmadash \prr{\Delta} : \t$, then $\Gamma \sigmadash \Delta\of\s\cap\t$, for some $\s$.

\item If $\Gamma \sigmadash \inll{\t}{\Delta} : \s\cup\t$, then $\Gamma \sigmadash \Delta : \s$ and $\Gamma\sigmadash \s\cup\t : \Type$.

\item If $\Gamma \sigmadash \inrr{\s}{\Delta} : \s\cup\t$, then $\Gamma \sigmadash \Delta : \t$ and $\Gamma\sigmadash \s\cup\t : \Type$.
\end{enumerate}
\end{lemma}


\begin{proposition}[Subderivation]\label{prp:subder:1} \hfill
\begin{enumerate}
\item A derivation of $\sigmadash \langle\rangle$ has a subderivation of\ $\Sigma\oksig$.
\item A derivation of $\Sigma, a\of K \oksig$ has subderivations of $\Sigma\oksig$ and $\sigmadash K$.
\item A derivation of $\Sigma, f \of \sigma\oksig$ has subderivations of $ \Sigma\oksig$ and $\sigmadash \sigma \of \Type$.
\item A derivation of $\sigmadash \Gamma, x \of \sigma$ has subderivations of $\Sigma\oksig$, $\sigmadash \Gamma$, and $\Gamma \sigmadash \sigma \of \Type$.
\item A derivation of $\Gamma \sigmadash \alpha$ has subderivations of $\Sigma\oksig$ and $\sigmadash \Gamma$.
\item Given a derivation of the judgement $\Gamma \sigmadash \alpha$, and a subterm occurring in the subject of this judgement, there exists a derivation of a judgement having this subterm as a subject.
\end{enumerate}
\end{proposition}


\addtocounter{theorem}{-14}

\begin{theorem}[Subject reduction of \DLF]\hfill 
\label{thm:subred}
\begin{enumerate}
\item If $\Gamma \sigmadash K$, and $K \to_{\Delta} K'$, then $\Gamma \sigmadash K'$.
\item If $\Gamma \sigmadash \sigma : K$, and $\sigma \to_{\Delta} \sigma'$, then $\Gamma \sigmadash \sigma' : K$.
\item If $\Gamma \sigmadash \Delta : \sigma$, and $\Delta \to_{\Delta} \Delta'$, then $\Gamma \sigmadash \Delta' : \sigma$.
\end{enumerate}
\end{theorem}

Finally, we define a possible algorithm for checking judgements in \DLF\ by computing a type or a kind for a term, and then testing for {\em definitional equality}, \ie\ $=_{\Delta}$, against the given type or kind. This is achieved by reducing both to their unique normal forms and checking that they are identical up to $\alpha$-conversion. Therefore we finally have:

\begin{theorem}[Decidability]
All the type judgments of \DLF\ are recursively decidable.
\end{theorem}

\addtocounter{theorem}{12}

\smallskip\noindent{\bf Minimal Relevant Implications and Type Inclusion.}
{Type inclusion} and the rules of {\em subtyping} are related to the notion of {minimal relevant implication}, see \cite{BM94,APLAS16}.
The insight is quite subtle, but ultimately very simple. This is what makes it appealing. The apparently intricate rules of subtyping and type inclusion, which occur in many systems, and might even appear {\em ad hoc} at times, can all be explained away in our principled approach, by proving that the relevant implication type is inhabited by a term whose essence is essentially a variable.

The following theorem we show how relevant implication subsumes the type-inclusion rules of the theory $\Xi$ of \cite{BDdL}, without rule (10): we call $\Xi'$ the resulting set. 
%
%
\addtocounter{theorem}{-12}
\begin{theorem}[Type Inclusion]
The judgement $\langle \rangle\sigmadash \Delta : \s\rightarrow^{\sf r} \t $ (where both $\s$ and $\t$ do not contain dependencies or relevant families) holds iff $\s \leq \t$ holds in the subtype theory $\Xi'$ of $\B$ enriched with new axioms of the form $\s_1\leq\s_2$ for each constant $c:\s_1\rightarrow^{\sf r}\s_2\in \Sigma$.

\begin{proof}
\newcommand{\coercion}[2]{\left\|#1\right\|_{#2}}

\hfill
\begin{itemize}

\item[] (if). Follows directly from Lemma \ref{prp:inversion}. 

\item[] (only if). It is possible to write a $\Delta$-term whose essence is an $\eta-$expansion of the identity $(\lambda x.x)$ corresponding to each of the axioms and rules in $\Xi'$. The $\Delta$-term is obtained by defining a function $\coercion{\s \leq \t}{\D}$, where $\s \leq \t$ is a subtyping derivation tree in the type theory $\Xi'$, which coerce a $\Delta$-term from type $\s$ to type $\t$:

\[
 \begin{array}[t]{l@{\quad}rcl}
 (1) & \coercion{\sigma \leqslant \sigma \cap \sigma}{\Delta} & \eqdef & \spair{\Delta}{\Delta} 
 \\[2mm]
 (2) & \coercion{\sigma \cup \sigma \leqslant \sigma}{\Delta} & \eqdef & \ssum{\lambda x\of\s.x}{\lambda x\of\s.x}\at \Delta 
 \\[2mm]
 (3) & \coercion{\sigma_1 \cap \sigma_2 \leqslant \sigma_i}{\Delta} & \eqdef & \pri \Delta 
 \\[2mm]
 (4) & \coercion{\sigma_i \leqslant \sigma_1 \cup \sigma_2}{\Delta} & \eqdef & \ini \Delta 
 \\[2mm]
 (6) & \coercion{\sigma \leqslant \sigma}{\Delta} & \eqdef & \Delta 
 \\[2mm]
 (7) & \coercion{\vcenter{\infer{\sigma_1 \cap \tau_1 \leqslant \sigma_2 \cap \tau_2}{\sigma_1 \leqslant \sigma_2 & \tau_1 \leqslant \tau_2}}}{\Delta} & \eqdef & \spair{\coercion{\sigma_1 \leqslant \sigma_2}{(\prl \Delta)}}{\coercion{\tau_1 \leqslant \tau_2}{(\prr \Delta)}} 
 \\[5mm]
 (8) & \coercion{\vcenter{\infer{\sigma_1 \cup \tau_1 \leqslant \sigma_2 \cup \tau_2}{\sigma_1 \leqslant \sigma_2 & \tau_1 \leqslant \tau_2 }}}{\Delta} & \eqdef & \ssum{\lambda x\of\sigma_1. \inll{\tau_2} \coercion{\sigma_1 \leqslant \sigma_2}{x}}{\lambda x\of\tau_1. \inrr{\sigma_2} \coercion{\tau_1 \leqslant \tau_2}{x}} \at \Delta
 \\[5mm]
 (9) & \coercion{\vcenter{\infer{\sigma \leqslant \rho}{\sigma \leqslant \tau & \tau \leqslant \rho}}}{\Delta} & \eqdef & \coercion{\tau \leqslant \rho}{(\coercion{\sigma \leqslant \tau}{\Delta})}
 \\[4mm]
 (11) & \coercion{(\sigma \rightarrow \tau) \cap (\sigma \rightarrow \rho) \leqslant \sigma \rightarrow (\tau \cap \rho)}{\Delta} & \eqdef & \lambda x \of \sigma. \spair{(\prl \Delta)\at x}{(\prr \Delta)\at x} 
 \\[2mm]
 (12) & \coercion{(\sigma \rightarrow \rho) \cap (\tau \rightarrow \rho) \leqslant (\sigma \cup \tau) \rightarrow \rho}{\Delta} & \eqdef & \lambda x\of\sigma\cup\tau. \ssum{\lambda y\of\sigma. (\prl \Delta)\at y}{\lambda y\of\tau. (\prr \Delta)\at y} \at x 
 \\[2mm]
 (14) & \coercion{\vcenter{\infer{\sigma_1 \rightarrow \tau_1 \leqslant \sigma_2 \rightarrow \tau_2}{\sigma_2 \leqslant \sigma_1 & \tau_1 \leqslant \tau_2}}}{\Delta} & \eqdef & \lambda x\of \sigma_2. \coercion{\tau_1\leqslant\tau_2}{(\Delta \at \coercion{\sigma_2\leqslant\sigma_1}{x})}
 \end{array}
 \]

 \end{itemize}

\end{proof}
\end{theorem}


%
%

As far as the $\lambda^{\Pi\&}$ system of Refinement Types introduced by Pfenning in~\cite{Refine93}, the following corollary holds:

\begin{corollary}[Pfenning's Refinement Types]\label{th:RefTyp}
The judgment $\vdash_\Sigma \sigma\leq\tau$ in $\lambda^{\Pi\&}$ can be encoded in \DLF\ by adding a constant of type $\sigma\rightarrow^{\sf r}\tau$ to $\Sigma'$, where the latter is the signature obtained from $\Sigma$ by replacing each clause of the form $a_1::a_2$ or $a_1\leq a_2$ in $\Sigma$ by a constant of type $a_1\rightarrow^{\sf r} a_2$.
\end{corollary}

\addtocounter{theorem}{10}

Moreover, while Pfenning needs to add explicitly the rules of subtyping (\ie\ the theory of $\leq$) in $\lambda^{\Pi\&}$, we inherit them naturally in \DLF\ from the rules for minimal relevant implication.

\newpage
\begin{landscape}
\subsection{Typed derivation of Pierce's example of Subsection 2.1}
\bigskip
\bigskip
 \[\
 \infer[(Conv)]{
 \Gamma \sigmadash
 \ssum{\lambda x_1\of\s_1.(\prl x) \at x_1 \at x_1)}{\lambda x_2\of\s_2.(\prr x) \at x_2 \at x_2)}
 \at ((\lambda x_3\of\rho \rightarrow \s_1 \cup \s_2.x_3) \at y \at z) : a\at (y \at z) \at (y \at z)}{
 \infer[(\Pi E)]{
 \Gamma \sigmadash \ssum{(\lambda x_1\of\s_1.(\prl x) \at x_1 \at x_1)}{(\lambda x_2\of\s_2.(\prr x) \at x_2 \at x_2)} \at ((\lambda x_3\of\rho \rightarrow \s_1 \cup \s_2.x_3) \at y \at z) : a\at ((\lambda x_3\of\rho \rightarrow \s_1 \cup \s_2.x_3) \at y \at z) \at ((\lambda x_3\of\rho \rightarrow \s_1 \cup \s_2.x_3) \at y \at z)}
 {
 \infer[(\cup E)]{\Gamma\sigmadash\ssum{\lambda x_1\of\s_1.(\prl x) \at x_1 \at x_1}{\lambda x_2\of\s_2.(\prr x) \at x_2 \at x_2} : \Pi x_4 \of \s_1\cup\s_2. a\at x_4 \at x_4}
 {
 \begin{array}{lr}
 \Gamma\sigmadash\lambda x_1\of\s_1.(\prl x) \at x_1 \at x_1 : \Pi x_1\of\s_1. 
 (a \at x_4 \at x_4) [\inll{\s_2}{x_1}/x_4]
 \\[1mm]
 \Gamma\sigmadash\lambda x_2\of\s_2.(\prr x) \at x_2 \at x_2 : \Pi x_2\of\s_2. 
 (a \at x_4 \at x_4)[\inrr{\s_1}{x_2}/x_4]
 \\[1mm]
 \Gamma, x_4\of\s_1\cup\s_2\sigmadash a\at x_4 \at x_4 : \Type
 \\[1mm]
 \essence{\lambda x_1\of\s_1.(\prl x) \at x_1 \at x_1} \mathrel{=_\eta} \essence{\lambda x_2\of\s_2.(\prr x) \at x_2 \at x_2)} \\
 \end{array}
 }
 &
 \Gamma\sigmadash (\lambda x_3\of\rho \rightarrow \s_1 \cup \s_2.x_3) \at y \at z : \sigma_1 \cup \sigma_2
 }
 }
\]

\bigskip 
\noindent where $$\Gamma \eqdef x \of \Pi x_1\of\s_1. \Pi x_2\of \s_1. a \at (\inll{\s_2} \at x_1) \at (\inll{\s_2} \at x_2) \cap \Pi x_1\of\s_2.\Pi x_2\of \s_2. a \at (\inrr{\s_1} \at x_1) \at (\inrr{\s_1} \at x_2), y \of \rho \rightarrow \s_1 \cup \s_2,z\of \rho$$ and $$\Sigma \eqdef a\of \s_1 \cup \s_2 \rightarrow \s_1 \cup \s_2 \rightarrow \Type$$

\end{landscape}

\end{document}